\documentclass[twocolumn,secnumarabic,amssymb, nobibnotes, aps, prl, showpacs]{revtex4-1}

\usepackage{graphicx}
\usepackage{dcolumn}
\usepackage{bm}

\begin{document}

\preprint{APS/123-QED}

\title{Charge-spin correlation in van der Waals antiferromagenet NiPS$_{3}$}

\author{So Yeun Kim,$^{1,2}$ Tae Yun Kim,$^{2,3}$ Luke J. Sandilands,$^{1,2}$ Soobin Sinn,$^{1,2}$ Min-Cheol Lee,$^{1,2}$ Jaeseok Son,$^{1,2}$ Sungmin Lee,$^{1,2}$ Ki-Young Choi,$^{1,2}$ Wondong Kim,$^{4}$ Byeong-Gyu Park,$^{5}$ C. Jeon,$^{6}$ Hyeong-Do Kim,$^{1,2}$ Cheol-Hwan Park,$^{2,3}$ Je-Geun Park,$^{1,2\ddag}$ S. J. Moon,$^{7}$* and T. W. Noh,$^{1,2\dag}$} 

\affiliation{$^{1}$Center for Correlated Electron Systems, Institute for Basic Science (IBS), Seoul 08826, Republic of Korea
\\$^{2}$Department of Physics \& Astronomy, Seoul National University (SNU), Seoul 08826, Republic of Korea
\\$^{3}$Center for Theoretical Physics, SNU, Seoul 08826, Republic of Korea
\\$^{4}$Korea Research Institute of Standards and Science (KRISS), Daejeon 34113, Republic of Korea
\\$^{5}$Pohang Accelerator Laboratory, Pohang University of Science and Technology (POSTECH), Pohang 37673, Republic of Korea
\\$^{6}$Advanced Nano-Surface Group, Korea Basic Science Institute (KBSI), Daejeon 34133, Republic of Korea
\\$^{7}$Department of Physics, Hanyang University, Seoul 04763, Republic of Korea}

\begin{abstract}
Strong charge-spin coupling is found in a layered transition-metal trichalcogenide NiPS$_{3}$, a van der Waals antiferromagnet, from our study of the electronic structure using several experimental and theoretical tools: spectroscopic ellipsometry, x-ray absorption and photoemission spectroscopy, and density-functional calculations. NiPS$_{3}$ displays an anomalous shift in the optical spectral weight at the magnetic ordering temperature, reflecting a strong coupling between the electronic and magnetic structures. X-ray absorption, photoemission and optical spectra support a self-doped ground state in NiPS$_{3}$. Our work demonstrates that layered transition-metal trichalcogenide magnets are a useful candidate for the study of correlated-electron physics in two-dimensional magnetic material.
\end{abstract}


\maketitle


It is a fundamental question of modern condensed matter physics how the three-dimensional (3D) physics that we are familiar with breaks down upon reducing the dimension. It is generally believed that in one-dimensional physical systems, dimensional fluctuations will become stronger and ultimately destroy any long-range order. However, the same question for two dimensions is nontrivial and extremely hard to solve.

For this goal of new physics, layered van der Waals (vdW) materials, such as graphene and transition metal (TM) dichalcogenides, have attracted much attention over the last decade \cite{1, 2, 3}. A remarkable advantage of these materials is that they can easily be mechanically exfoliated to produce two-dimensional (2D) crystals \cite{1, 2, 3}. The intriguing collective quantum phenomena found in 2D vdW materials include charge density waves and superconductivity \cite{4, 5, 6, 7, 8}. These observations open a new window of opportunity for novel device applications through the manipulation of collective quantum states in atomically thin 2D phases \cite{3, 6, 9}. Despite the extensive study of vdW materials, it is striking that there have been very few studies of magnetic 2D vdW materials.

Only recently has attention been focused on new magnetic 2D vdW materials of ternary transition-metal trichalcogenide (TMTC) families, such as Cr\textit{B}Te$_{3}$ (\textit{B} = Si or Ge) and \textit{TM}PX$_{3}$ (\textit{TM} = 3\emph{d} TMs; \textit{X} = chalcogens) \cite{10, 11, 12}. These TMTC bulk samples exhibit various magnetic orderings: ferromagnetic (FM), zig-zag antiferromagnetic (AF), N\'{e}el AF, and stripy AF \cite{11,12}. Moreover, all three key magnetic Hamiltonians, i.e., Ising, XY, and Heisenberg types, are realized in \textit{TM}P\textit{X}$_{3}$ \cite{13, 14, 15, 16}. Of further interest, long-range magnetic ordering has recently been reported to persist even in the monolayer limit, e.g., FePS$_{3}$ and CrGeTe$_{3}$ \cite{10, 10-1, 14}. Thus, it has become clear that the TMTC families, i.e., the layered vdW magnets, are a fertile playground for exploring rich and intriguing phenomena related to 2D magnetism, which will ultimately pave the way to novel applications in spintronics.

Compared with other nonmagnetic vdW materials, these new magnetic vdW materials offer another distinct and fundamental opportunity, namely insight into strongly correlated electronic systems in the 2D limit. Over the past few decades, researchers have attempted to unravel correlation physics using 2D materials \cite{17, 18, 19}; e.g., oxide ultrathin films produced by pulsed laser deposition (PLD) and molecular beam epitaxy (MBE), where a variety of emergent phenomena are reported to take place \cite{21, 22}. With magnetism being an intrinsic property of TMTC, it is natural for us to expect to see some correlation physics, even in the 2D limit of those materials. It is well-known that in magnetic TM oxides, electron-electron correlation plays a central role in magnetic interactions, including superexchange and double exchange mechanisms \cite{25,24}. For instance, many 3\textit{d} TM oxides exhibit AF ordering as a consequence of the superexchange processes associated with \textit{d}-\textit{d} hopping, whose amplitude is directly related to the correlation interaction energy. To date, the electron correlation effects in TMTC remain largely unexplored experimentally, despite their potential importance.

In this Letter, we report on the electronic structure of bulk NiPS$_{3}$ single crystal, a TMTC antiferromagnet in which the electronic and spin (or magnetic) structures are closely related to one another. Using optical spectroscopy techniques we observed clear anomalies in the optical spectral weight at the N\'{e}el temperature, driven mainly by the magnetic ordering, which is a hallmark of correlated electronic systems \cite{26,27}. Subsequent x-ray absorption and photoemission studies also showed, together with cluster model calculations, that NiPS$_{3}$ is a self-doped negative charge transfer (NCT) insulator. That is, the ligand (sulfur atoms) has a strong hole-like character due to a NCT energy, unlike more conventional systems. Such an intriguing electronic and magnetic ground state was also confirmed by density functional theory (DFT) with Coulomb interaction (\textit{U}) calculations. Our studies identify NiPS$_{3}$ as a unique vdW magnet showing clear evidence of strong correlation and NCT behavior.

\begin{figure}
	\includegraphics[width=8cm]{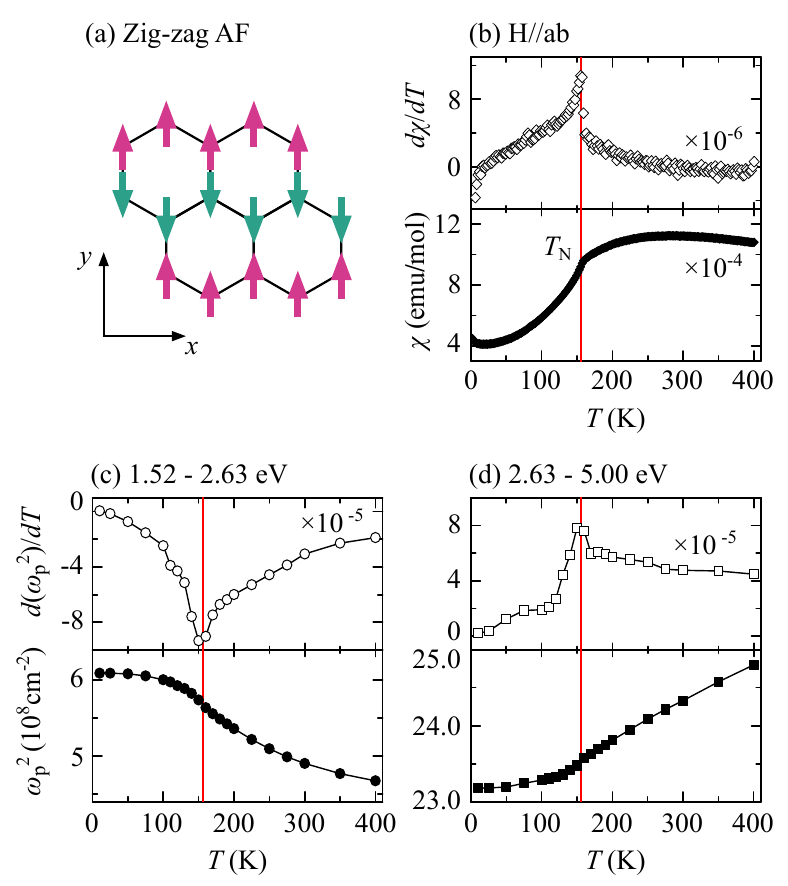}
	\caption{(color online). (a) Schematic diagram of the zig-zag antiferromagnetic ordering. The honeycomb lattice of Ni ions is shown by black lines, with the spin depicted as an arrow in each site. (b) The magnetic susceptibility ($\chi$) of NiPS$_{3}$ measured under in-plane bias fields (lower panel) and its first derivative with respect to \textit{T} (upper panel). (c)\textendash(d) The spectral weight ${\omega}_{p}^{2}$ and its first derivative (upper panel) obtained for the energy range of (c) 1.52\textendash2.62 eV and (d) 2.63\textendash5.00 eV. The N\'{e}el temperature of NiPS$_{3}$ ($T_{N}$ $\sim$ 154 K) is represented as a red vertical line in panels (b)\textendash(d).}
\end{figure}

High-quality single crystals of NiPS$_{3}$ were grown by a chemical vapor transport method, as described in Ref. \cite{28, 28-1}. For the optical measurements, we used a 55-$\mu$m-thick single crystal sample and an M-2000 ellipsometer (J. A. Woollam Co.). In addition, we carried out the DFT+\textit{U} calculations using the Quantum Espresso package \cite{31, 32, 33, 34, 35, 36, 37, 38} (for details on the DFT method, see Supplemental Materials (a-1) \cite{30}).

NiPS$_{3}$ is one of the transition-metal phosphorous trichalcogenides (\textit{TM}P\textit{X}$_3$) with an AF long-range ordering at 154 K. It has a monoclinic structure with C$^{2}_{2h}$  symmetry and features edge-sharing NiS$_6$ octahedra arranged on a honeycomb lattice \cite{28,39}. At the center of the honeycomb lattice, two P atoms are located above and below the TM plane. They are then connected to the S atoms like a dumbbell, forming a (P$^2$S$^6$)$^{4\textendash}$ bipyramid structure \cite{39,40}. It is known that the magnetic moments of Ni ions are aligned in a so-called “zig-zag” pattern; i.e., chains of ferromagnetically coupled spins are arranged antiferromagnetically, as shown in Fig. 1(a) \cite{32}. Our in-plane magnetic susceptibility $\chi$(\textit{T}) shows an AF magnetic anomaly at the N\'{e}el temperature ($T_{N}$ $\sim$ 154 K) [Fig. 1(b)].

The optical conductivity spectra $\sigma_{1}(\omega$) show that NiPS$_{3}$ is an insulator with an optical gap of about 1.8 eV [Fig. 2(a)]. At the same time, there are strong narrow absorption peaks in the visible-ultraviolet energy regions, near 2.2, 3.5, and 4.6 eV [labeled A, B, and C, respectively, in Fig. 2(a)]. Below the gap, there exists two additional weak peaks near 1.1 and 1.7 eV denoted as $\alpha$ and $\beta$, respectively [see the inset of Fig. 2(a)]. For comparison, NiO, which has the same formal valence of Ni (+2) as NiPS$_{3}$, is reported to have weak on-site \textit{d}-\textit{d} transitions at 1.13 and 1.75 eV \cite{41}. Due to the similarities in the energy positions and strengths of the peaks, it is most likely that the $\alpha$ and $\beta$ peaks are the on-site \textit{d}-\textit{d} transitions, a transition within one Ni ion. It should be noted that the hybridization of Ni 3\textit{d} orbitals with S 3\textit{p} can provide a much larger bandwidth than that with O 2\textit{p} in NiO. Nonetheless, the narrowness of the A, B, and C peaks, in addition to the existence of the $\alpha$ and $\beta$ peaks, suggest the presence of well-localized Ni 3\textit{d} orbital states near the Fermi surface (\textit{E}$_F$) in the electronic structure.

\begin{figure}
	\includegraphics[width=8cm]{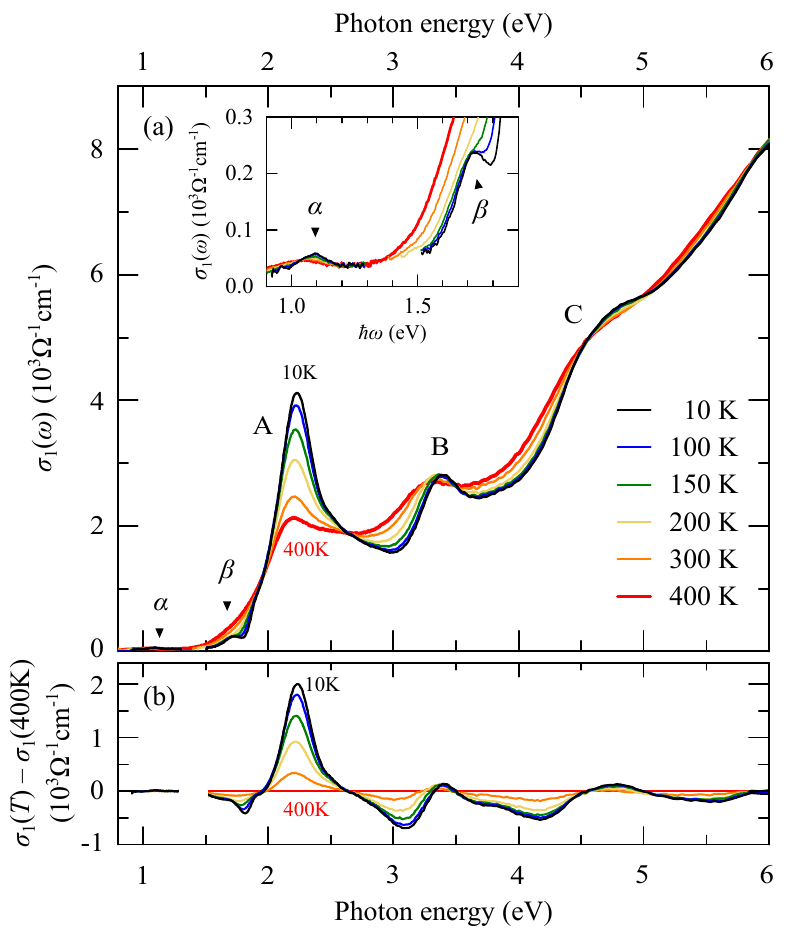}
	\caption{(color online). (a) Real part of the optical conductivity $\sigma_{1}(\omega$) of NiPS$_{3}$ measured between 10 and 400 K. The three main optical transitions are labeled A, B, and C. Two weak on-site \textit{d}-\textit{d} transitions ($\alpha$ and $\beta$) are located below 2 eV. The inset shows the $\alpha$ and $\beta$ transitions in detail. (b) Difference spectra $\sigma_{1}(\omega, \textit{T}$)\textendash $\sigma_{1}(\omega, 400 K$). Fabry-P\'{e}rot interference fringes are observed in a narrow energy range near 1.4 eV and below 0.9 eV, where the sample becomes transparent; these are not shown.}
\end{figure}

When we measure the temperature dependence of the optical conductivity of NiPS$_{3}$, it displays strong temperature variations. For example, Fig. 2(b) shows the difference curve of  $\sigma_{1}(\omega$) at various temperatures after subtracting off the 400 K spectrum. Note that peak A shows strong enhancement in intensity while peaks B and C sharpen without much change in intensity upon cooling. To quantify the temperature evolution, we calculated the squares of the plasma frequency $\omega_{p}^{2}$: $\omega_{p}^{2}(T)$ = $8\int\sigma_{1}(\omega, T)d\omega$ \cite{42}, which we refer to as the spectral weight (\textit{SW}). The optical sum rule is satisfied when we integrate up to 4.2 eV, implying the clear \textit{SW} shifts from peak A to peaks B and C (see Supplemental Materials (b) \cite{30}). At the same time, satisfaction of the optical sum rule indicates that most of the temperature-dependent changes in the electronic structure should occur near the Fermi level, \textit{E}$_{F}$.

When we further examined the temperature dependence, we found clear anomalies in the \textit{SW} changes of the main peaks occurring at \textit{T}$_N$. Figures 1(c) and 1(d) show the \textit{SW} changes in the energy ranges of 1.52\textendash2.63 eV and of 2.63\textendash5.00 eV, respectively. The former represents the \textit{SW} changes of peak A, and the latter represents those of peaks B and C. The lower panel of Fig. 1(c) shows that the ${\omega}_{p}^{2}$(\textit{T})  of peak A increased with decreasing temperature, with a kink structure at \textit{E}$_{F}$. The anomaly was seen more clearly in the first derivative of ${\omega}_{p}^{2}$, as plotted on the upper panel. A similar feature was also observed in the temperature evolution of the \textit{SW} of peaks B and C, as shown in Fig. 1(d). Thus, the observation of \textit{SW} anomalies at \textit{E}$_{F}$ constitutes the compelling evidence of strong coupling between the electronic and spin structures in NiPS$_{3}$.

Conventionally, most ground states of strongly correlated compounds can be categorized via the Zaanen-Sawatzky-Allen (ZSA) classification scheme \cite{43,44}. Depending on the relative size of on-site \textit{U} and charge-transfer energy ($\Delta$), the insulating ground state can be classified as either Mott-Hubbard or charge transfer insulators. In the former, the lowest energy charge fluctuation occurs between inter-site TM ions (\textit{d}$^{n}$\textit{d}$^{n}$) $\,\to\,$ (\textit{d}$^{n+1}$\textit{d}$^{n-1}$) with an energy cost of \textit{U}, and in the latter, between ligand and TM ions (\textit{d}$^{n}$\textit{p}$^{6}$ $\,\to\,$ \textit{d}$^{n+1}$\textit{p}$^{5}$) with $\Delta$ (see Supplemental Materials (d) \cite{30}). However, there is a less-explored region in the ZSA scheme, where compounds can have NCT energy ($\Delta$ $\textless$ 0). In this region, electrons of the ligand \textit{p}-orbital transfer to the TM \textit{d} levels in the ground state, creating holes at the ligands without external doping. Because of this, they are often referred to as ``self-doped'' \cite{45}. The ligand hole can naturally make an important contribution to the conductivity and magnetism \cite{46}. To some extent, the nature of the self-doped state is analogous to the ground state of the Zhang-Rice singlet states found in hole-doped cuprates \cite{47,48}.

Indeed, this interpretation of the self-doped ground state agrees with XAS and XPS results of NiPS$_{3}$. We obtained the Ni \textit{L}$_{2,3}$ edge XAS spectra of bulk NiPS$_{3}$ which was subsequently analyzed using a cluster model (see Supplemental Materials (e-2) \cite{30}). At the \textit{L}$_{3}$ edge, NiPS$_{3}$ show a main peak near 851 eV, which are clearly distinguished from that of compounds with large positive $\Delta$ values, such as NiO ($\Delta \sim$ 4.6 eV) or NiCl$_{2}$ ($\Delta \sim$ 3.6 eV), where the main peak shows clear splitting \cite{49}. We simulated the Ni \textit{L}$_{2,3}$ edge XAS spectra using cluster model and found the NiPS$_{3}$ belongs to the regime of $\Delta \leq$ 0 eV (see Supplemental Materials (e-3) [30]). In addition, the Ni 2\textit{p}$_{3/2}$ core XPS spectrum is found to be nearly identical to that of NiGa$_{2}$S$_{4}$, a first reported NCT sulfide \cite{50, 55} (see Supplemental Materials (e-2) \cite{30}). The NiS$_{6}$ cluster model used on NiGa$_{2}$S$_{4}$ also reproduces the Ni 2\textit{p}$_{3/2}$ XPS spectrum of NiPS$_{3}$ well, using the parameters of $\Delta$ = \textendash1.0 and \textit{U} = 5.0 eV. Note the calculated ground state is given as $\Psi_{g} = {\alpha}\left|d^{8}\right\rangle + {\beta}\left|d^{9}\underline{\textit{L}}\right\rangle + {\gamma}\left|d^{10}\underline{\textit{L}}^{2}\right\rangle$, where \underline{\textit{L}} indicates a ligand (sulfur) hole, with ${\alpha}^2$ = 0.25, ${\beta}^2$ = 0.60 and ${\gamma}^2$ = 0.15. Above findings support NiPS$_{3}$ should have a self-doped ground state with dominant $d^{9}\underline{\textit{L}}$ character, and demonstrate NiPS$_{3}$ as the only known example of a magnetically ordered vdW material with an NCT insulating state.

To gain further insight, we performed DFT+\textit{U} calculations. We used a magnetic ground state with a zig-zag AF ordering [Fig. 1(a)] and an effective \textit{U} value of 4 eV. The solid line in Fig. 3(a) shows the theoretical calculation of in-plane $\sigma_{1}(\omega$) after rescaling the data, in a way that satisfies the sum rule, to match the energy position of the first transition peak (see Supplemental Materials (a-2) \cite{30}). The theoretical $\sigma_{1}(\omega$) reproduced most of the key features found in our experimental data.

\begin{figure}
	\includegraphics[width=8cm]{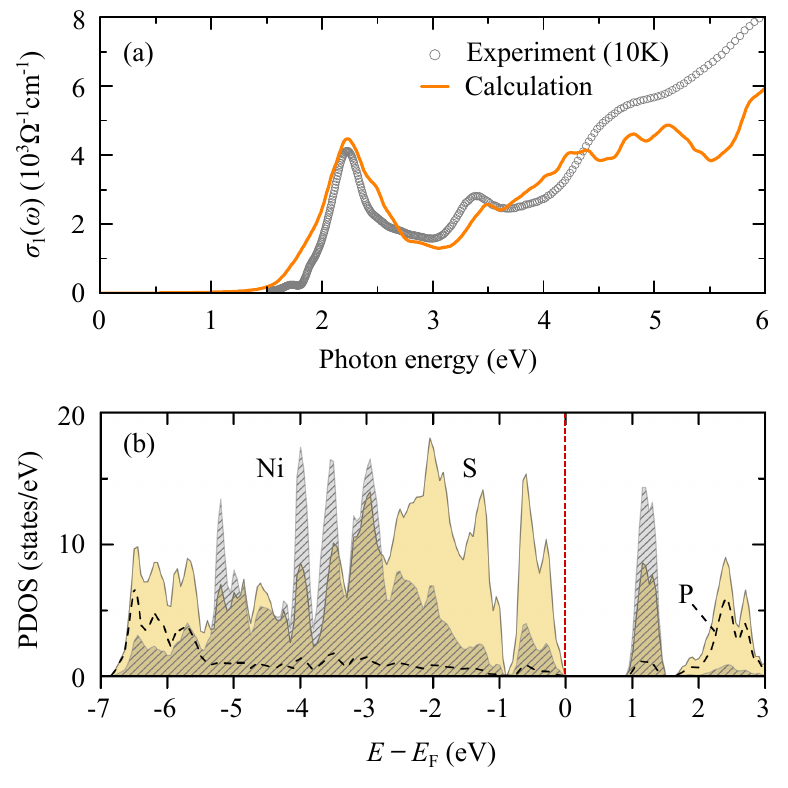}
	\caption{(color online). (a) Real part of the optical conductivity $\sigma_{1}(\omega$) obtained from experiments conducted at 10 K (circle) and from DFT+\textit{U} after a stretch in the energy axis by 1.39 and a proper renormalization satisfying the sum rules \cite{28}. (b) Projected density of states (PDOS) for Ni (gray oblique lines), S (yellow), and P (dashed line) obtained from DFT+\textit{U} calculations.}
\end{figure}

The DFT+\textit{U} calculations also support our interpretation of the self-doped NCT ground state for NiPS$_{3}$. Figure 3(b) shows the projected density of states (PDOS) for Ni, P, and S orbitals in NiPS$_{3}$; as shown by the dashed curve, the P 3\textit{p} states are located mostly above 2 eV or below \textendash5 eV, and show a strong hybridization with the S 3\textit{p} orbitals as a result of covalent bonding in the P$_{2}$S$_{6}$ bipyramids \cite{39,40}. On the other hand, the PDOS for Ni and S orbitals is much higher near \textit{E}$_{F}$ than that of P 3\textit{p} [Fig. 3(b)]. Thus, the valence bands near \textit{E}$_{F}$ are mainly S 3\textit{p} orbital states, and the occupied Ni 3\textit{d} orbital states are located mostly at lower energies. Narrow Ni 3\textit{d} bands hybridized with the S 3\textit{p} show up at $\sim$1.3 eV above \textit{E}$_{F}$. The S 3\textit{p} orbitals are located at higher energies than most Ni \textit{d} orbitals, which indicates charge transfer from S to Ni ions \textit{in the ground state}, and is consistent with the NCT picture \cite{46}. Thus, the ground state is expected to have strong contributions from $d^{9}\underline{\textit{L}}$  and $d^{10}\underline{\textit{L}}^{2}$ configurations, in addition to $d^{8}$.

The total number of \textit{d}-electrons from the PDOS analysis of our DFT+\textit{U} was larger than expected from the formal valence, which is consistent with a self-doped NCT ground state. While Ni$^{2+}$ ion in the formal valence is expected to have an occupation number of 8, we obtained 8.6 from our PDOS analysis. Such an increase in the occupation number was also found in a DFT study of Cs$_{2}$Au$_{2}$Cl$_{6}$, which showed a larger occupation number of \textit{d}-electrons than expected from the formal valence because $\Delta$ $\textless$ 0 \cite{56}.

The picture of the self-doped ground state also finds support from the magnetic properties. According to our DFT+\textit{U} calculations, the theoretical value of the ordered magnetic moment is 1.24 ${\mu}_B$ per Ni ions, as opposed to the formal value of 2 ${\mu}_B$. This reduction in the magnetic moment can be easily understood using the results of our cluster calculations. As mentioned earlier, the ground state consists of three configurations, each with different spin numbers: $d^{8}$ (\textit{S}$_{Ni}$ = 1), $d^{9}\underline{\textit{L}}$ (\textit{S}$_{Ni}$ = 1/2), and $d^{10}\underline{\textit{L}}^{2}$ (\textit{S}$_{Ni}$ = 0). In addition, our calculations show that four S ions in each NiS$_{6}$ octahedron have spin polarizations (0.15 ${\mu}_B$) aligned in the same direction as the spin of the central Ni ion. The remaining S positioned in the middle of two adjacent FM chains, as well as the P ions, are nonmagnetic (see Supplemental Materials (a-3) \cite{30}). Here, we would like to point out that a recent neutron diffraction experiment on NiPS$_{3}$ also reported a large reduction in the Ni moment with an experimental ordered moment of 1.05 ${\mu}_B$ \cite{32}. We speculate that this reduction in the magnetic moment of Ni ions, consistently supported by our theory and experiment, is a natural consequence of the charge transfer energy being negative.

Let us revisit the temperature-dependent \textit{SW} anomaly, shown in Figs. 1(c) and 1(d), within the simple cluster picture. In the $d^{9}\underline{\textit{L}}$ state, the lowest energy transition, peak A, is expected to be enhanced when the neighboring Ni ions are bonded antiferromagnetically. The corresponding ground state should be composed of \textit{t}$_{2g}$$^6$\textit{e}$_{g}$$^3$ electrons at Ni 3\textit{d} orbitals and one hole at S 3\textit{p} orbitals ($\underline{\textit{L}}$); i.e., \textit{t}$_{2g}$$^6$\textit{e}$_{g}$$^3$$\underline{\textit{L}}$, with the $^3$\textit{A}$_{2g}$ symmetry \cite{50} (see Supplemental Materials (f) \cite {30}). According to the NiS$_6$ cluster model, the lowest-energy ionization state was found to have $^2$\textit{E}$_{g}$ symmetry \cite{50, 57}, and the corresponding state is clearly seen in XPS valence spectra (see Supplemental Materials (e-4) \cite {30}). This suggests that peak A could be assigned as an inter-site transition between NiS$_6$ clusters that transfer an electron between two \textit{t}$_{2g}$$^6$\textit{e}$_{g}$$^3$$\underline{\textit{L}}$ ($^3$\textit{A}$_{2g}$) clusters. These transitions would split the ground state into a low spin \textit{t}$_{2g}$$^6$\textit{e}$_{g}$$^2$$\underline{\textit{L}}$/\textit{t}$_{2g}$$^6$\textit{e}$_{g}$$^3$$\underline{\textit{L}}$$^2$ state on one site and a low spin \textit{t}$_{2g}$$^6$\textit{e}$_{g}$$^3$/\textit{t}$_{2g}$$^6$\textit{e}$_{g}$$^4$$\underline{\textit{L}}$ state on the other site. Considering spin conservation, such transitions may be allowed between AF-bonded clusters, but forbidden between FM-bonded clusters.

Therefore, the temperature-dependent \textit{SW} and its anomaly should arise from the increased AF bonds in zig-zag-ordered honeycomb lattice. At the temperature above \textit{T}$_{N}$, there will be no preference between AF and FM bonds of NiS$_{6}$ clusters. Below \textit{T}$_{N}$, however, the number of AF and FM bonds will differ: For the nearest neighboring Ni ions (with exchange interaction \textit{J}$_{1}$), there will be 2 FM and a AF bonds; for the second-nearest (\textit{J}$_{2}$), 2 FM and 4 AF bonds; and for the third-nearest (\textit{J}$_{3}$), 3 AF bonds only. Recent calculations on the monolayer of the NiPS$_{3}$ predicted that \textit{J}$_{3}$ is nearly four times larger than \textit{J}$_{1}$, while the value of \textit{J}$_{2}$ is very small \cite{58, 59}. Likewise, the dominant strength of \textit{J}$_{3}$ is consistent with the increased number of AF bonds below \textit{T}$_{N}$, and consequently the increased intensity of peak A. The \textit{SW}, therefore, should reflect the degree of AF correlation present at a given temperature, which is consistent with \textit{SW} anomalies observed across \textit{T}$_{N}$ in the experiments [Fig. 1(c)]. 

In summary, we investigated the electronic structures of NiPS$_{3}$, one of the layered vdW antiferromagnets. Our results showed strong charge-spin coupling; i.e., a close relationship between the electronic structure and the magnetic ordering. We also found that NiPS$_{3}$ is a rare self-doped NCT insulator; i.e., a strong hole-character is present in the ground state. Our findings provide a basis to explore a novel material class of vdW magnets, which could lead to rich and intriguing phenomena related to 2D magnetism and enable new applications in novel spintronic devices.

\begin{acknowledgements}
We gratefully acknowledge insightful discussions with K. W. Kim, K. Burch and B. C. Park. This work was supported by Institute for Basic Science (IBS) in Korea (Grant No. IBS-R009-D1, IBS-R009-G1). S. J. Moon was supported by Basic Science Research Program through the National Research Foundation of Korea (NRF) funded by the Ministry of Science, ICT and Future Planning (Grant No. 2014R1A2A1A11054351 and 2017R1A2B4009413). T. Y. Kim and C. -H. Park were supported by Korean NRF-2016R1A1A1A05919979. Experiments at PLS-II were supported in part by MSIP and POSTECH. The STXM experiment was supported by the National Council of Science \& Technology grant (Grant No. CAP-16-01-KIST and NRF-2016K1A3A7A09005335). W. Kim specially thanks to N. Kim and H. Shin for their supporting for STXM measurement at PLS-II.

\end{acknowledgements}

\textit{E-mail address}:

*soonjmoon@hanyang.ac.kr

$^{\dag}$twnoh@snu.ac.kr

$^{\ddag}$jgpark10@snu.ac.kr


\begin{references}
\bibitem{28-1} We subsequently checked the stoichiometry using a scanning electron microscope (COXI EM30; COXEM), equipped with an energy-dispersive x-ray spectrometer (Bruker Quantax 100), which was followed by magnetic susceptibility measurements obtained with a SQUID-VSM (MPMS 3; Quantum Design).
\bibitem{55} It seems that the lower electronegativity of chalcogen ions in TMTC can also drive the system into NCT regime, as in the cases of some TM oxides \cite{45, 44, 46, 47}.
\end{references}
\end{document}